\documentclass{aa}  

\usepackage{graphicx}
\usepackage{txfonts}
\usepackage{xcolor}
\usepackage{siunitx}
\usepackage{multirow}
\usepackage{lineno}

\begin{document}

   \title{New evidence supporting past dust ejections from active asteroid (4015) Wilson--Harrington}

   \author{Sunho Jin,
          \inst{1,2}
          Masateru Ishiguro,
          \inst{1,2}
          Jooyeon Geem,
          \inst{1,2}
          Hiroyuki Naito,
          \inst{3}
          Jun Takahashi,
          \inst{4}
          Hiroshi Akitaya,
          \inst{5,6,7}
          Daisuke Kuroda,
          \inst{8}
          Seitaro Urakawa,
          \inst{8}
          Seiko Takagi,
          \inst{9}
          Tatsuharu Oono,
          \inst{10}
          Tomohiko Sekiguchi,
          \inst{11}
          Davide Perna,
          \inst{12}
          Simone Ieva,
          \inst{12}
          Yoonsoo P. Bach,
          \inst{13}
          Ryo Imazawa,
          \inst{7,14}
          Koji S. Kawabata,
          \inst{7,14}
          Makoto Watanabe,
          \inst{15}
          \and
          Hangbin Jo
          \inst{1,2}
          }

   \institute{Department of Physics and Astronomy, Seoul National University, 1 Gwanak-ro, Gwanak-gu, Seoul 08826, Republic of Korea
   \and SNU Astronomy Research Center, Department of Physics and Astronomy, Seoul National University, 1 Gwanak-ro, Gwanak-gu, Seoul 08826, Republic of Korea \and Nayoro Observatory, 157-1 Nisshin, Nayoro, Hokkaido, 096-0066, Japan \and Center for Astronomy, University of Hyogo, 407-2 Nishigaichi, Sayo, Hyogo 679-5313, Japan \and Astronomical Research Center, Chiba Institute of Technology, 2-17-1 Tsudanuma, Narashino, Chiba 275-0016, Japan \and Planetary Exploration Research Center, Chiba Institute of Technology, 2-17-1 Tsudanuma, Narashino, Chiba 275-0016, Japan \and Hiroshima Astrophysical Science Center, Hiroshima University, 1-3-1 Kagamiyama, Higashi-Hiroshima, Hiroshima, 739-8526, Japan \and Bisei Spaceguard Center, Japan Spaceguard Association, 1716-3 Okura, Bisei-cho, Ibara, Okayama 714-1411, Japan \and Department of Earth and Planetary Sciences, Faculty of Science, Hokkaido University, Kita-ku, Sapporo, Hokkaido 060-0810, Japan \and Department of Cosmosciences, Graduate School of Science, Hokkaido University, Kita-ku, Sapporo, Hokkaido 060-0810, Japan \and Asahikawa Campus, Hokkaido University of Education, Hokumon, Asahikawa, Hokkaido 070-8621, Japan \and INAF - Osservatorio Astronomico di Roma, Via Frascati 33, I-00078, Monte Porzio Catone, Italy \and Korea Astronomy and Space Science Institute (KASI), 776 Daedeok-daero, Yuseong-gu, Daejeon, 34055, Republic of Korea \and Physics Program, Graduate School of Advanced Science and Engineering, Hiroshima University, 1-3-1 Kagamiyama, Higashi-Hiroshima, Hiroshima 739-8526, Japan \and Department of Physics, Okayama University of Science, 1-1 Ridai-cho, Kita-ku, Okayama, Okayama 700-0005, Japan\\
              \email{jsh854@snu.ac.kr, ishiguro@snu.ac.kr}
             }

   \date{Received ; accepted }

 
  \abstract
   {(4015) Wilson--Harrington (hereafter, WH) was discovered as a comet in 1949 but has a dynamical property consistent with that of a near-Earth asteroid. Although there is a report that the 1949 activity is associated with an ion tail, the cause of the activity has not yet been identified.}
   {This work aims to reveal the mysterious comet-like activity of the near-Earth asteroid.}
   {We conducted new polarimetric observations of WH from May 2022 to January 2023, reanalyses of the photographic plate images taken at the time of its discovery in 1949, and dust tail simulation modelings, where the dust terminal velocity and ejection epoch are taken into account.}
   {We found that this object shows polarization characteristics similar to those of low-albedo asteroids. We derived the geometric albedo ranging from $p_V = 0.076 \pm 0.010$ to $p_V = 0.094 \pm 0.018$ from our polarimetry (the values vary depending on the data used for fitting and the slope-albedo relationship coefficients). In addition, the 1949 image showed an increase in brightness around the nucleus. Furthermore, we found that the color of the tail is consistent with sunlight, suggesting that the 1949 activity is associated with dust ejection. From the dust tail analysis, $\sim 9 \times 10^5$ kg of material was ejected episodically at a low velocity equivalent to or even slower than the escape velocity.}
   {We conclude that WH is most likely an active asteroid of main belt origin and that the activity in 1949 was likely triggered by mass shedding due to fast rotation.}

               
               \titlerunning{Evidence of past dust ejections from (4015) Wilson--Harrington}
               \authorrunning{Jin et al.}

   \maketitle
%

\section{Introduction} \label{sec:intro}
Traditionally, small Solar System bodies (SSSBs) in the inner region have been thought to be distinctively classified as either comets or asteroids. In that case, comets are icy bodies that formed beyond the snow line and display comae and tails near perihelia due to the volatile sublimation. Meanwhile, asteroids are rocky objects that originate from the inner part of the Solar System and appear as point sources in the observed images. However, discoveries have expanded this traditional classification to a broader continuum, with comets and asteroids as the end members \citep{Jewitt2022}. Active asteroids, whose orbits and spectra are similar to those of asteroids but display comet-like activity, are a kind of object between comets and asteroids \citep{Jewitt2012}.

107P/(4015) Wilson--Harrington (hereafter, WH) is an intriguing object that lies between comets and active asteroids. In November 1949, when it was discovered, WH displayed a tail in photographic plate images \citep{Cunningham1950}, but never again \citep{Cunningham1950, Bowell1992, Ishiguro2011}. Accordingly, it was classified as a dormant comet whose activity was extinct \citep[e.g.,][]{Chamberlin1996, Coradini1997, Lupishko2001}. \citet{Fernandez1997} conducted Finson-Probstein modeling on the tail of WH and found that the dust size would be tens to hundreds of micrometers, which contradicts the blue color ($B - R \sim -1$) of the tail reported by \citet{Bowell1992}. Thus, they suggested that the observed tail is dust-less and due to $\mathrm{CO}^+$ and $\mathrm{H}_2\mathrm{O}^+$ ion fluorescence, which are commonly found in comets. To date, it remains the only explanation for the activity of WH.

As \citet{Fernandez1997} pointed out, however, the existence of an ion tail without the presence of dust is questionable if this object was almost a dormant comet. This is because the surface of the object would be covered with a dust mantle, and the drag force of sublimating gas from beneath the mantle would naturally carry out dust particles from the object. In addition, the spectrum indicated an asteroidal rather than cometary \citep[i.e., C-complex asteroids or carbonaceous chondrites,][]{Tholen1984, Kareta2023}. Moreover, \citet{Bottke2002} suggested from their dynamical simulation that its origin is more likely from the outer main belt (65 \%) rather than the Jupiter family comet (4 \%). This implies that the object could be an active asteroid that originated from the main asteroid belt. If this is indeed an active asteroid, several dust ejection mechanisms, such as impacts or rotational breakups, found in several active asteroids, could also be possible triggers for the activity \citep{Jewitt2012}.

In this study, we aim to determine whether WH is a dormant comet or an active asteroid and to constrain its activity mechanism. First, we conducted the first polarimetric study on the object and compared its surface scattering properties with those of comets and asteroids. Polarimetry is a unique tool for discriminating between asteroid-like and comet-like objects, as demonstrated by \citet{Geem2022}. Moreover, polarization degrees obtained at high phase angles (i.e., Sun-object-observer's angle $\alpha > 40\degr$) of the active asteroid (3200) Phaethon are significantly higher than those of a comet nucleus, further suggesting the usability of polarimetry at large phase angles \citep{Ito2018, Kuroda2022}. Thus, it is safe to say that polarimetry provides critical information for distinguishing between comets and asteroids. Next, we reanalyzed the photographic plates of WH in 1949 to inspect the color of its tail. Besides, We conducted a dust ejection simulation to determine physical quantities related to its activity, following the method in \citet{Ishiguro2007}. Considering all this information (i.e., nuclear polarimetry, dust color, and dust ejection mechanism), we provide a comprehensive discussion about a possible mechanism for the 1949's activation.


\section{Methods} \label{sec:methods}
\subsection{Polarimetry} \label{subsec:pol}

\begin{table*}
\caption{Summary of polarimetric observation}             
\label{table:1}      
\centering  

\begin{tabular}{cccccccccc}      
\hline\hline       
Date & UT & Telescopes/Instruments & Exptime \tablefootmark{a} (s) & N \tablefootmark{b} & Airmass & r \tablefootmark{c} (au) & $\Delta$ \tablefootmark{d} (au) & $\alpha$ \tablefootmark{e} ($\degr$) & $\phi$ \tablefootmark{f} ($\degr$)\\ 
\hline                    
2022-05-25 & 16:17--17:35 &   Pirka/MSI &  420 & 12 & 2.4--3.6 & 1.51 &   0.84 &  39.24 & 252.70 \\
2022-06-06 & 16:48--17:16 &   Pirka/MSI &  300 &  8 & 2.2--2.4 & 1.41 &   0.69 &  42.37 & 249.92 \\
2022-06-14 & 16:00--16:53 &   Pirka/MSI &  300 & 12 & 2.1--2.8 & 1.34 &   0.60 &  45.27 & 248.13 \\
2022-07-01 & 15:52--17:10 &   Pirka/MSI &  180 & 24 & 1.6--2.3 & 1.21 &   0.46 &  54.90 & 245.61 \\
2022-07-09 & 14:33--15:18 &   Pirka/MSI &  180 & 16 & 2.7--4.1 & 1.15 &   0.42 &  61.13 & 245.94 \\
2022-07-10 & 15:13--17:10 &   Pirka/MSI &  180 & 36 & 1.5--2.8 & 1.14 &   0.41 &  62.00 & 246.09 \\
2022-07-21 & 17:00--17:26 &   Pirka/MSI &  180 & 12 & 1.4--1.6 & 1.07 &   0.40 &  71.09 & 249.46 \\
2022-07-26 & 15:46--15:46 &   Pirka/MSI &  240 &  4 & 2.3--2.3 & 1.05 &   0.40 &  74.56 & 251.90 \\
2022-07-29 & 15:35--16:23 &   Pirka/MSI &  240 & 16 & 1.9--2.5 & 1.03 &   0.41 &  76.32 & 253.55 \\
2022-07-31 & 16:34--16:50 &   Pirka/MSI &  240 &  8 & 1.7--1.8 & 1.02 &   0.42 &  77.34 & 254.72 \\
2022-08-07 & 17:09--17:09 &   Pirka/MSI &  180 &  4 & 1.6--1.6 & 0.99 &   0.44 &  79.74 & 258.90 \\
2022-08-10 & 16:18--17:35 &   Pirka/MSI &  180 & 20 & 1.5--2.1 & 0.99 &   0.46 &  80.24 & 260.66 \\
2022-08-14 & 15:48--16:01 &   Pirka/MSI &  180 &  8 & 2.4--2.6 & 0.98 &   0.48 &  80.44 & 262.92 \\
2022-08-14 & 18:29--18:51 &  Nayuta/WFGS2 &  300 &  8 & 1.4--1.5 & 0.98 &   0.48 &  80.40 & 262.92\\
2022-09-14 & 17:39--18:32 &   Pirka/MSI &  240 & 12 & 1.4--1.6 & 1.01 &   0.65 &  70.77 & 275.77 \\
2022-12-05 & 17:31--19:20 & Nayuta/WFGS2 &  300 & 24 & 1.0--1.1 & 1.62 &   0.79 &  27.04 & 283.43 \\
2022-12-21 & 14:31--15:00 &   Pirka/MSI &  180 & 12 & 1.2--1.2 & 1.76 &   0.82 &  14.57 & 283.68 \\
2023-01-12 & 17:34--18:26 &   Pirka/MSI &  180 & 12 & 1.5--1.8 & 1.94 &   0.96 &   2.74 &  73.48 \\
2023-01-25 & 11:00--17:10 & Kanata/HONIR &  120 & 88 & 1.0--1.5 & 2.05 &   1.10 &  10.36 &  89.74 \\
\hline                  
\end{tabular}
\tablefoot{
\tablefoottext{a}{Exposure time in seconds}
\tablefoottext{b}{Number of valid exposures}
\tablefoottext{c}{Median heliocentric distance in au}
\tablefoottext{d}{Median geocentric distance in au}
\tablefoottext{e}{Median solar phase angle in degrees}
\tablefoottext{f}{Position angle of the scattering plane in the equatorial coordinate system (J2000) in degrees.}
}
\end{table*}

Table \ref{table:1} provides a summary of our polarimetric observations of WH. We conducted polarimetric observations from May 25, 2022 to January 25, 2023 using three telescopes in Japan: the 1.6-m Pirka telescope at the Nayoro Observatory (NO) of the Faculty of Science, Hokkaido University (Minor Planet Center observatory code Q33), the 2-m Nayuta telescope at the Nishi-Harima Astronomical Observatory (NHAO), operated by the University of Hyogo\footnote{Latitude: 35 01 31 N, Longitude: 134 20 08 E, Elevation: 449m}, and the 1.5-m Kanata telescope at the Higashi-Hiroshima Observatory (HHO)\footnote{Latitude: 34 22 39 N, Longitude: 132 46 36 E, Elevation: 503m}. The observations covered a wide phase angle range, from 2.7$\degr$ to 80.4$\degr$. At NO, we utilized the polarimetry mode of the Multi-Spectral Imager (MSI), which has a pixel scale of 0.39$\arcsec$ pixel$^{-1}$ \citep{Watanabe2012}. Meanwhile, at the NHAO, we employed the polarimetry mode of the Wide Field Grism Spectrograph 2 (WFGS2), which has a pixel scale of 0.198$\arcsec$ pixel$^{-1}$ \citep{Uehara2004, Kawakami2021}, and at the HHO, we used the Hiroshima Optical and Near-InfraRed camera (HONIR), which has a pixel scale of 0.29$\arcsec$ pixel$^{-1}$ \citep{Akitaya2014}. We used a standard R$_{\mathrm{C}}$-band filter and set an exposure time range of 60 to 180 seconds to obtain polarimetric accuracy of less than 1\%p in non-sidereal tracking mode. The observations were conducted for 16 nights at NO, 2 nights at NHAO, and one night at HHO.

We conducted our data analysis using the same methods as those described in detail in \citet{Ishiguro2022} for MSI data and in \citet{Geem2022b} for WFGS2 and HONIR data. The analysis process can be outlined as follows:
\begin{enumerate}
\item preprocessing of the data,
\item masking of the field stars near the target and bad pixels, 
\item aperture photometry to derive the source flux of ordinary and extraordinary components,
\item derivation of the Stokes parameters ($Q/I$, and $U/I$),
\item correction of the polarization efficiency, instrumental polarization, and position angle offset, and
\item derivation of the polarization degree ($P$), the position angle ($\theta_\mathrm{P}$), the polarization degree with respect to the scattering plane normal($P_\mathrm{r}$), and the position angle with respect to the scattering plane normal($\theta_\mathrm{r}$).
\end{enumerate}

    \begin{figure*}
    \centering
    \includegraphics[width=\hsize]{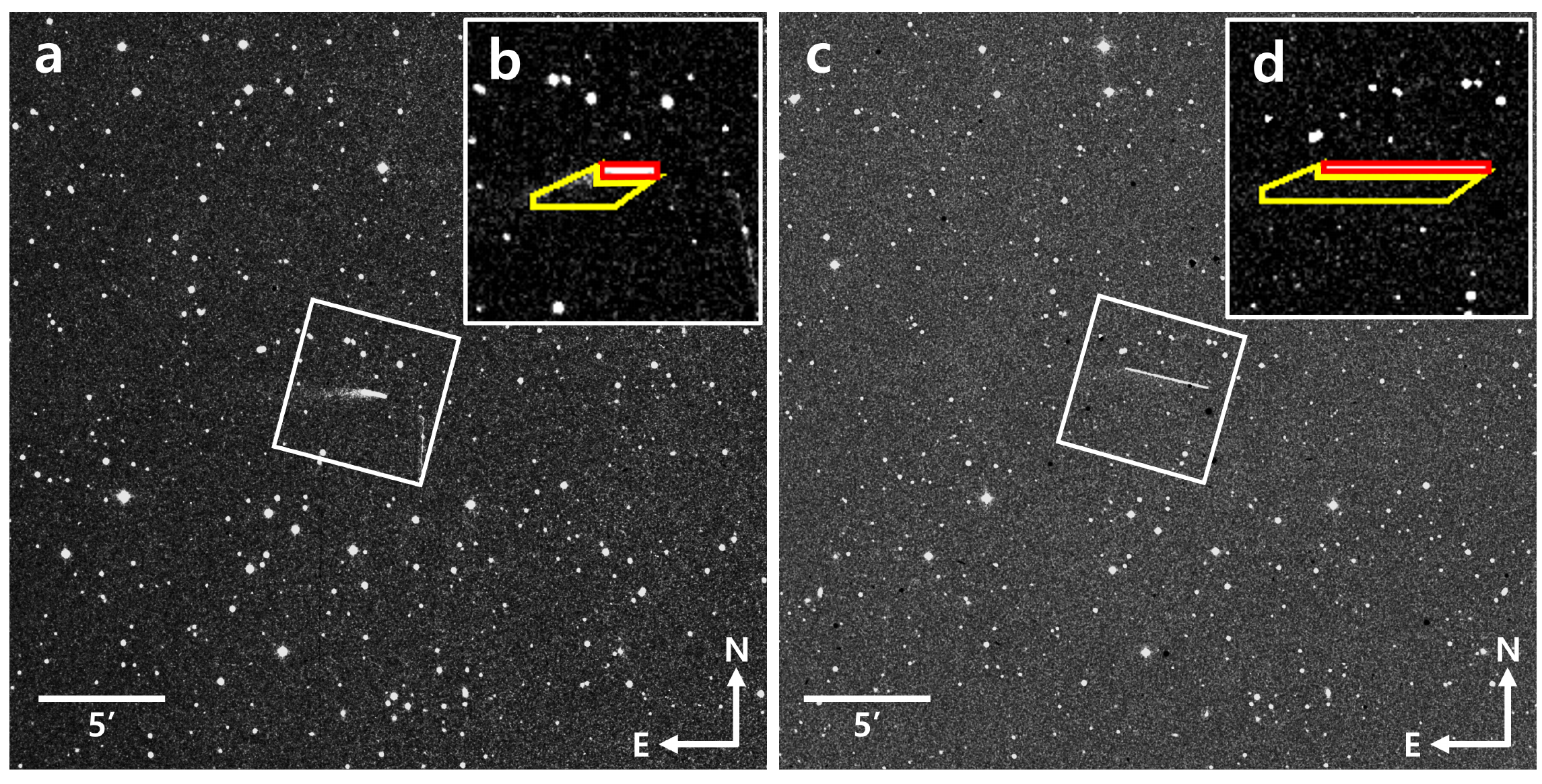}
        \caption{Digitized photographic plate images of WH in 1949 used for the analysis. (a) and (c) are the blue and red plate images taken with exposure times of 720 seconds and 2700 seconds, respectively. In these images, the nucleus and tails are stretched out due to the sidereal tracking mode of the telescope mount. The tail in the red plate image is not as clear as the one in the blue plate because of the lower sensitivity of the red plate. (b) and (d) are zoomed images within the white squares of images (a) and (c). These images are rotated to align the WH's velocity vectors in a horizontal direction. The areas surrounded by red rectangles indicate the region for the nucleus photometry, while those surrounded by the lower left parts of the red rectangles and yellow lines indicate the region for the tail photometry.
              }
        \label{Fig1}
   \end{figure*}

\subsection{Photometry}\label{subsec:photo}
We obtained two digitized photographic plate images of WH in 1949 from the Mikulski Archive for Space Telescopes (MAST) Digitized Sky Survey (DSS) \footnote{https://stdatu.stsci.edu/cgi-bin/dss\_form}. One was taken with a photographic plate with an emulsion 103aO, sensitive to blue wavelengths (hereafter, blue plate), and the other was with an emulsion 103aE, sensitive to red wavelengths (red plate). These plates were taken on November 19, 1949, during the first Palomar Observatory Sky Survey (POSS). The coordinates of the image centers are taken from Table 1 of \citet{Fernandez1997}, and the size of each image is $30\arcmin \times 30\arcmin$. The pixel sizes of those images are 15.0 \textmu m (1.0 \arcsec) and 25.284 \textmu m (1.67 \arcsec) for blue and red plates, respectively.

Files downloaded from the MAST contain the photographic emulsion density multiplied by a constant factor (6\,553.4) to adjust the 16-bit representation \citep{booklet_n}. We linearized these emulsion densities to intensities following the technique in \citet{Cutri1993}. Detailed procedures are given in Appendix A. Then, we subtracted the background using Source-Extractor \citep{Bertin1996}. We set the mesh size for the background estimation to be 512 pixels, large enough not to be significantly affected by the diffuse tail of WH. Moreover, we estimated the error of these intensities using the method described in Appendix B. The error estimated in Appendix B does not include the inherent root mean square (RMS) error of diffuse source photometry from the POSS plate. \citet{Cutri1993} suggested the amount of the RMS error to be 0.1 to 0.3. Therefore, we added the RMS error of 0.2 to our original results.

Because of the sidereal tracking, the nucleus appeared elongated in the images. We determined regions where the WH nucleus and tail were located from the images (see Fig. \ref{Fig1}). We visually determined the rectangular regions and hexagonal regions that enclose the nucleus and the tails in each band for the photometry. The leftmost and rightmost sides of the tail area have relative position angles of 157$\degr$ and 144$\degr$ with respect to the horizontal axis of each image (i.e., the WH's velocity vector projected on the sky plane). We obtained these angles from Fig. 9 of \citet{Fernandez1997}. The distance between the lower side of the nucleus area and the tail area corresponds to 42$\arcsec$ on both plates, although the horizontal lengths are different due to the different exposure times. As a result, the areas of the tail regions are 3\,834.8 $\mathrm{arcsec}^2$ and 8\,292.2 $\mathrm{arcsec}^2$ for blue and red plates, respectively.

We calculated the intensities and uncertainties of the nucleus and tail within these regions for blue and red plates. For the tail, which is an extended source, we derived the surface brightness by dividing the intensity by the photometric areas. Once the intensities and surface brightness were obtained, we estimated the mass of the dust tail ($M_{\mathrm{tail}}$). It can be given as

\begin{equation}
    M_{\mathrm{tail}} = \frac{4\pi N_0 \rho}{3\left(4+q\right)} 
    \left[a^{4+q}_\mathrm{max}-a^{4+q}_\mathrm{min}\right] ~,
    \label{Eq:mass}
\end{equation}

\noindent where $\rho$ is the dust mass density. We assumed $\rho = 1300$ kg m$^{-3}$ based on the measurement of Ryugu samples due to the similarity of the spectral type (i.e., C-complex). We supposed a simple power law size distribution of the dust particles in the radius range from 
$a_\mathrm{min}$ to $a_\mathrm{max}$. $q$ denotes the power index of the differential size distribution, and $N_0$ is the reference number of dust particles with the radius $a_0$ = 1 m. Assuming the scattering property (i.e., the albedo and the scattering phase functions) of dust particles are the same as that of the nucleus, we derived $N_0$ as below:


\begin{equation}
    N_0 = \frac{R^2_{\mathrm{WH}}\left(3+q\right)}{a^{3+q}_\mathrm{max}-a^{3+q}_\mathrm{min}} \left(\frac{I_{\mathrm{tail}}}{I_{\mathrm{WH}}}\right)~, 
    \label{Eq:No}
\end{equation}

\noindent where $I_\mathrm{WH}$ and $I_\mathrm{tail}$ are intensities in the blue plate within the enclosed areas after sky subtractions. We cited the radius of WH nucleus ($R_\mathrm{WH}$) to be 2\,200 m (i.e., the best-fit value obtained in \citealt{Bach2017}).

\subsection{Dust ejection simulation}\label{subsec:tailmodel}
Assuming that the tail is associated with scattered light by ejected dust, we analyzed the tail properties. We conducted a simple simulation of the dust ejection to estimate the ejection epoch and particle size. The model descriptions are given in, for example, \citet{Ishiguro2007}. 

First, we produced the synchrone--syndyne network, where the zero ejection speed was considered. The synchrone curves are the lines indicating the distribution of dust particles ejected at given epochs. On the other hand, the syndyne curves represent the location of dust particles with the same $\beta$ value, a ratio between solar radiation pressure ($F_\mathrm{r}$) and gravity ($F_\mathrm{g}$). It is given as

\begin{equation}
    \beta \equiv \frac{F_\mathrm{r}}{F_\mathrm{g}}=\frac{3 L_{\odot} Q_{\mathrm{pr}}}{16\pi G M_{\odot}c\rho a}=\frac{K Q_{\mathrm{pr}}}{\rho a}~,
\end{equation}

\noindent where $L_{\odot}$, $G$, $M_{\odot}$, and $c$ are the solar luminosity, gravitational constant, solar mass, and light speed, respectively. Substituting these constant values, $K$ is equal to $5.7\times10^{-4}$ kg m$^{-2}$. $Q_{\mathrm{pr}}$ is the radiation pressure coefficient averaged over the solar spectrum. We considered large absorbing particles and assumed $Q_{\mathrm{pr}} = 1$, following \citet{Ishiguro2007}. The synchrone--syndyne network was drawn on the $XY$-plane of the digitized blue photographic plate image taken on November 19, 1949.
Furthermore, we conducted a three-dimensional analysis of the dust tail, allowing consideration of the terminal ejection velocity of the particles, described in \citet{Ishiguro2007}. This model assumes a terminal ejection velocity ($v_{\mathrm{ej}}$) of particles as

\begin{equation}
    v_{\mathrm{ej}} = V_0 \beta^{u_1} \left(\frac{r_\mathrm{h}}{1 \mathrm{au}}\right)^{-u_2} ,
        \label{Eq:V0}
\end{equation}

\noindent where $V_0$ is the reference ejection velocity (m s$^{-1}$) of the particles with $\beta=1$ at the heliocentric distance of $r_\mathrm{h}$=1 au. We set $u_1$ and $u_2$ as 0.5. With the given $v_{\mathrm{ej}}$ and $\beta$, we solved Kepler's equation to determine the spatial distribution of dust particles. Moreover, we assumed the differential size distribution of dust particles in the size range between $a_{\textrm{min}}$ and $a_{\textrm{max}}$ and the power index $q$.


%

%

\begin{table*}
\caption{Summary of nightly-averaged linear polarization degrees}             
\label{table:2}      
\centering  

\begin{tabular}{cccccccccc}      
\hline\hline   
      Date & UT &  Telescopes/Instruments &  $\alpha$\tablefootmark{a} (deg) &     P \tablefootmark{b} (\%) & $\sigma_P$ \tablefootmark{c} (\%) &  $\theta$ \tablefootmark{d} (deg) &  $\sigma_\theta$ \tablefootmark{e} (deg) & $P_r$ (\%) \tablefootmark{f} (\%) &  $\theta_r$ \tablefootmark{g} (deg)\\
\hline
2022-05-25 & 16:17--17:35 &  Pirka/MSI &  39.24 &  8.11 & 1.30 & -21.78 &    4.58 &  8.01 &  -184.47 \\
2022-06-06 & 16:48--17:16 &   Pirka/MSI &  42.37 & 11.05 & 1.06 & -25.17 &    2.74 & 10.88 &  -185.08 \\
2022-06-14 & 16:00--16:53 &   Pirka/MSI &  45.27 & 12.47 & 1.14 & -23.36 &    2.63 & 12.45 &  -181.48 \\
2022-07-01 & 15:52--17:10 &   Pirka/MSI &  54.90 & 16.42 & 0.21 & -23.23 &    0.36 & 16.41 &  -178.83 \\
2022-07-09 & 14:33--15:18 &   Pirka/MSI &  61.13 & 20.04 & 0.41 & -23.93 &    0.58 & 20.04 &  -179.87 \\
2022-07-10 & 15:13--17:10 &   Pirka/MSI &  62.00 & 21.08 & 0.19 & -22.74 &    0.26 & 21.06 &  -178.83 \\
2022-07-21 & 17:00--17:26 &   Pirka/MSI &  71.09 & 27.28 & 0.42 & -20.62 &    0.44 & 27.28 &  -180.09 \\
2022-07-26 & 15:46--15:46 &   Pirka/MSI &  74.56 & 27.00 & 1.92 & -12.79 &    2.04 & 26.53 &  -174.68 \\
2022-07-29 & 15:35--16:23 &   Pirka/MSI &  76.32 & 28.54 & 0.45 & -15.87 &    0.45 & 28.53 &  -179.42 \\
2022-07-31 & 16:34--16:50 &   Pirka/MSI &  77.34 & 30.77 & 0.60 & -14.50 &    0.56 & 30.76 &  -179.22 \\
2022-08-07 & 17:09--17:09 &   Pirka/MSI &  79.74 & 30.79 & 0.86 & -11.05 &    0.80 & 30.79 &  -179.95 \\
2022-08-10 & 16:18--17:35 &   Pirka/MSI &  80.24 & 32.24 & 0.62 &  -8.49 &    0.55 & 32.23 &  -179.16 \\
2022-08-14 & 15:48--16:01 &   Pirka/MSI &  80.44 & 35.72 & 1.49 &  -6.00 &    1.19 & 35.69 &  -178.92 \\
2022-08-14 & 18:29--18:51 &   Nayuta/WFGS2 &  80.4 & 33.26 & 3.79 &  4.39 &    4.29 & 30.66 &  -168.59 \\
2022-09-14 & 17:39--18:32 &   Pirka/MSI &  70.77 & 24.88 & 1.32 &  11.50 &    1.52 & 24.39 &     5.73 \\
2022-12-05 & 17:31--19:20 & Nayuta/WFGS2 &  27.04 &  2.13 & 0.33 &   8.70 &    4.39 &  2.10 &    -4.73 \\
2022-12-21 & 14:31--15:00 &   Pirka/MSI &  14.57 &  0.64 & 0.55 & -66.81 &   24.61 & -0.60 &    99.51 \\
2023-01-12 & 17:34--18:26 &   Pirka/MSI &   2.74 &  2.26 & 1.06 & -69.96 &   13.46 & -0.66 &   126.55 \\
2023-01-25 & 11:00--17:10 & Kanata/HONIR &  10.36 &  1.12 & 0.36 &  80.47 &    9.86 & -1.06 &    88.78 \\
\hline
\end{tabular}
		\tablefoot{		                     
			\tablefoottext{a}{Median solar phase angle in degrees,}
                \tablefoottext{b}{Nightly averaged linear polarization degree in percent,}
			\tablefoottext{c}{Error of P in percent,}
			\tablefoottext{d}{Direction of the semi-major axis of polarization ellipse with respect to the celestial north in the equatorial coordinate system (J2000) in degrees,}
			\tablefoottext{e}{Error of $\theta_P$ in degrees,}
			\tablefoottext{f}{Polarization degree referring to the scattering plane normal in percent,}
                \tablefoottext{g}{Direction of the semi-major axis of polarization ellipse with respect to the scattering plane normal in degrees}}

\end{table*}

    \begin{figure}
    \centering
    \includegraphics[width=\hsize]{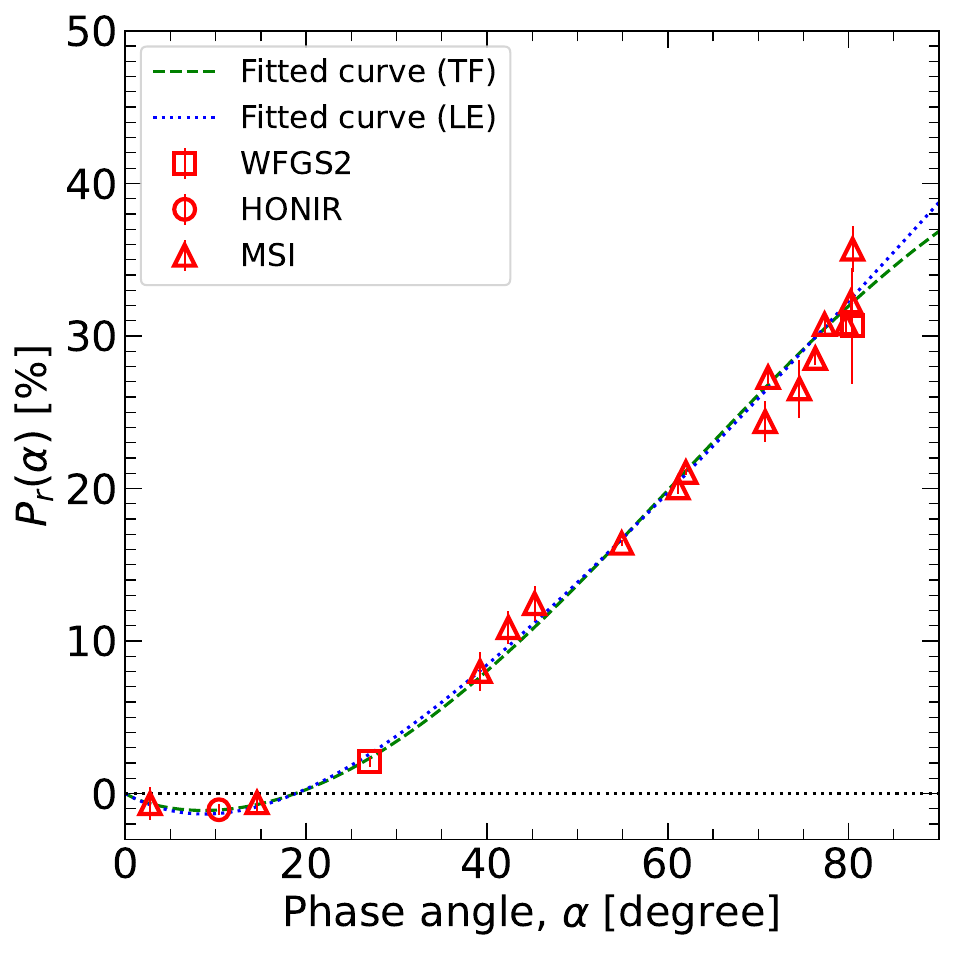}
        \caption{Polarization phase curve (PPC) of WH. Each symbol denotes observations from different observatories and instruments. A green dotted line shows a fitted line using a trigonometric function (TF, Eq. \ref{Eq:TR}), and a blue dashed line shows a fitted line using a linear-exponential function (LE, Eq. \ref{Eq:LE}).   
              }
        \label{Fig2}
   \end{figure}  

    \begin{figure}
    \centering
    \includegraphics[width=\hsize]{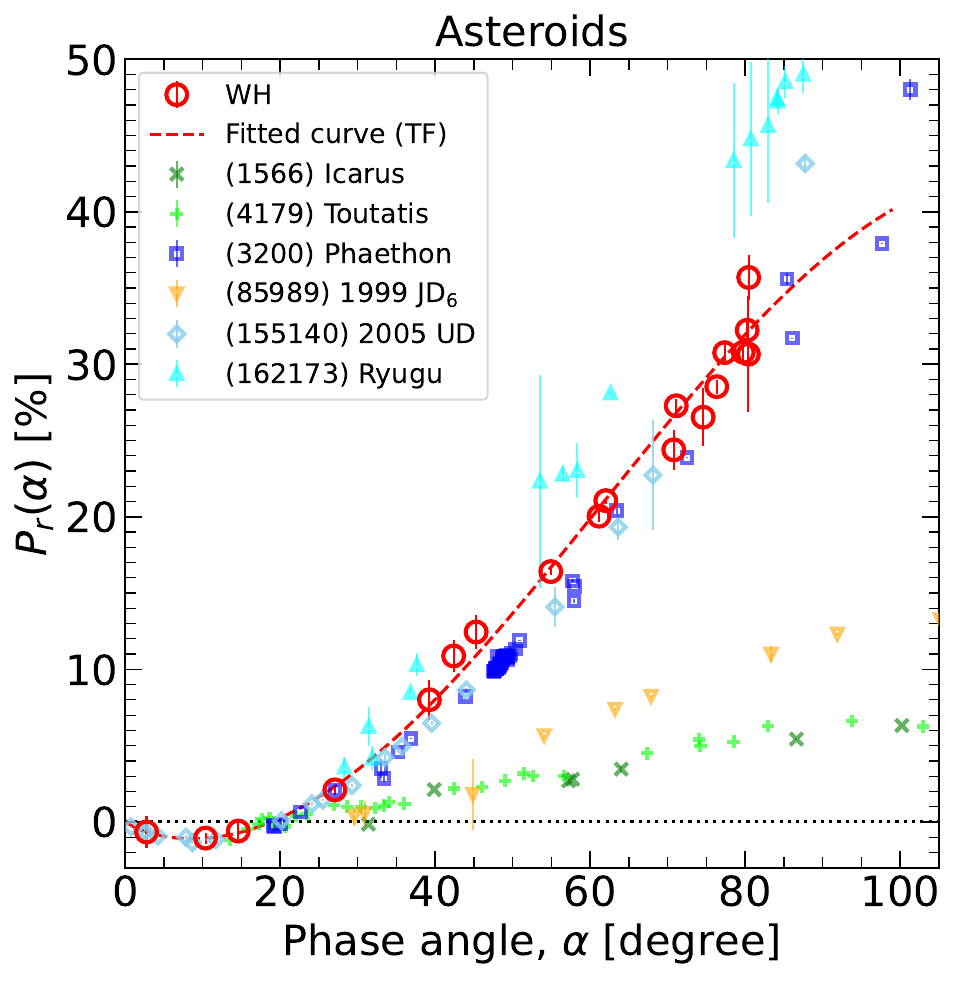}
        \caption{
        Comparison in polarization phase curves (PPCs) of WH (red circles) with fitted curve using Eq. \ref{Eq:TR} and other asteroids: two B-type asteroids (3200) Phaethon \citep[blue squares,][]{Ito2018, Shinnaka2018, Devogele2018}, (155140) 2005 UD \citep[cyan diamonds,][]{Ishiguro2022}, a C-type asteroid (162173) Ryugu \citep[cyan upper triangles,][]{Kuroda2022}, two S-type asteroids (1566) Icarus \citep[green crosses,][]{Ishiguro2017}, and (4179) Toutatis \citep[lime pluses,][]{Lupishko1995, Mukai1997}, and a L-type asteroid (85989) 1999 JD$_6$  \citep[orange lower triangles,][]{Kuroda2021}.  
              }
        \label{Fig3}
   \end{figure} 

    \begin{figure}
    \centering
    \includegraphics[width=\hsize]{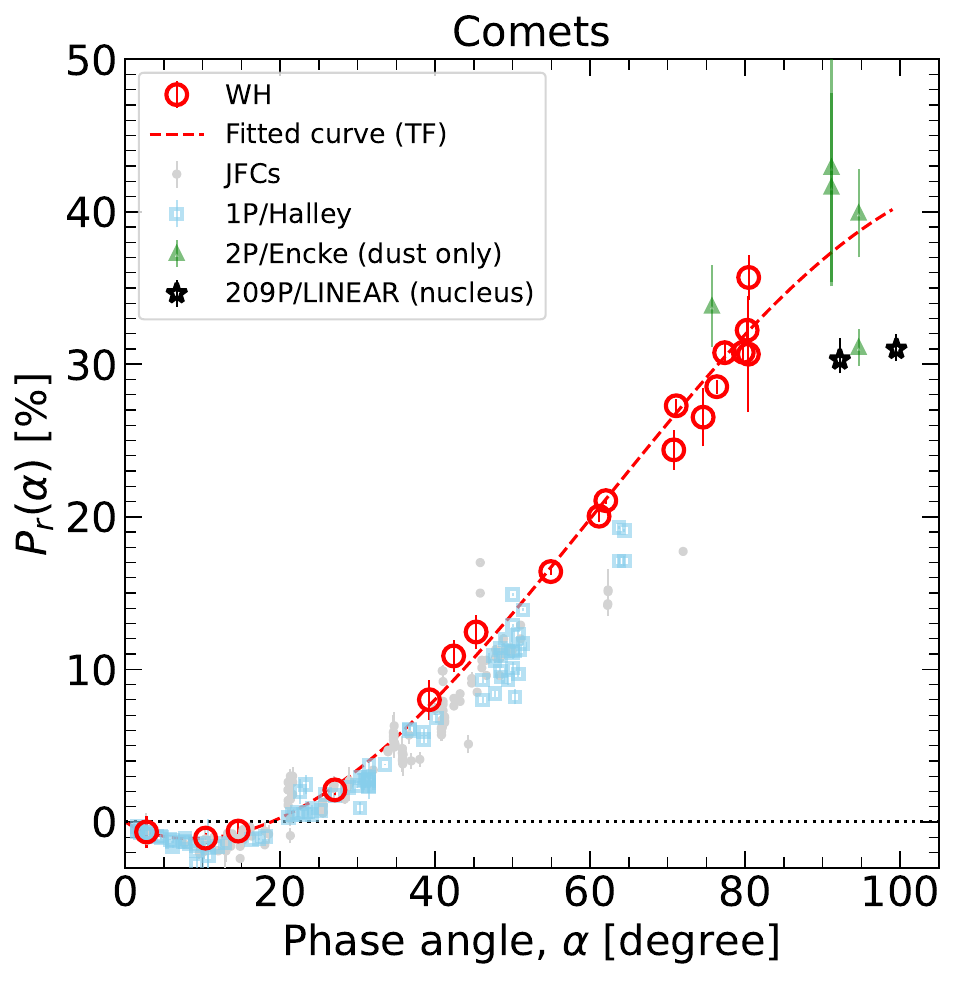}
        \caption{
        Comparison in polarization phase curves (PPCs) of WH (red circles)  with fitted curve using Eq. \ref{Eq:TR}, a nucleus of Jupiter family comet 209P/Linear \citep[black stars,][]{Kuroda2022}, dust particles in comae of Jupiter Family Comets \citep[grey circles,][]{Kiselev2005}, 1P/Halley \citep[cyan squares,][]{Kikuchi1987, Chernova1993}, and polarization degrees from dust continuum of 2P/Encke \citep[green upper triangles,][]{Jockers2005, Kwon2018}.
              }
        \label{Fig4}
   \end{figure}

\section{Results} \label {sec:res}
In this section, we report our polarimetric results in Sect. \ref{subsec:pol_res}, photometric results from the 1949 observation images in Sect. \ref{subsec:pho_res}, and the tail analysis in Sect. \ref{subsec:tail_res}, as described below.

\subsection{Polarimetric results} \label{subsec:pol_res}
Throughout our observations, we did not detect any signatures of comet-like activity (i.e., a tail and a coma) in our polarimetric images in 2022--2023. Therefore, we characterize the polarimetric properties of the nuclear surface.

We summarized nightly-averaged linear polarization degrees in Fig. \ref{Fig2} and Table \ref{table:2}. Similar to other SSSBs, it indicates the negative branch at $\alpha\lesssim 20\degr$ and the positive branch at $\alpha\gtrsim 20\degr$ \citep{Ishiguro2022}. 
Although we cannot determine the maximum polarization degree because of insufficient phase angle coverage, we find
$P_r(\alpha) > 30 \%$ at $\alpha \sim 80 \degr$.

We compared the polarization phase curve ($P_r(\alpha)$, PPC) of WH with other asteroids (Fig. \ref{Fig3}). At a glance, the PPC's slope near the inversion angle is predominantly steeper than that of S-complex asteroids and consistent with those of C-complex asteroids. A careful look reveals that the WH's PPC is more moderate than that of Ryugu and steeper than that of Phaethon around the low phase angles. This fact suggests that the WH's geometric albedo value would be between these two asteroids. We also compared WH's PPC with those of a comet nucleus (209P/LINEAR) and cometary dust (Fig. \ref{Fig4}). Although there is only one comet whose nuclear PPC is available, we find that WH's PPC indicates a higher polarization degree than 209P. Considering most SSSBs have polarization maxima around $\alpha>90\degr$ while WH was observed at $\alpha<<90\degr$ but indicated the large polarization degree ($P_r(80\degr)>$30 \%), this object likely shows a polarization maximum larger than that of comet nuclei.

To determine the polarimetric parameters quantitatively, we employed PyMC3 \citep{Salvatier2016} to fit the PPC using the empirical trigonometric function (TF) shown in \citet{Ishiguro2022}, modified from \citet{Lumme1993} as shown below:

\begin{equation}
    P_r\left(\alpha\right) = h \left(\frac{\sin \alpha}{\sin \alpha_0}\right)^{c_1} \left(\frac{\cos \frac{\alpha}{2}}{\cos\frac{\alpha_0}{2}}\right)^{c_2} \sin\left(\alpha-\alpha_0\right) ,
    \label{Eq:TR}
\end{equation}

\noindent where $\alpha_0$ is an inversion angle where $P_r(\alpha_0) = 0$ \% is satisfied, and $h$ is a polarimetric slope at $\alpha=\alpha_0$. $c_1$ and $c_2$ are constant values obtained from fitting. Meanwhile, We also fitted PPC using a modified linear-exponential function (LE) from \citet{Bach2024} modified from the original forms in \citet{2002MmSAI..73..716M, 2003Icar..161...34K},

\begin{equation}
    P_r\left(\alpha\right) = h \frac{(1-e^{-\alpha_0/k})\alpha - (1-e^{-\alpha_0/k})\alpha_0}
    {1-(1+\alpha_0/k)e^{-\alpha_0/k}} ,
    \label{Eq:LE}
\end{equation} 

\noindent such that

\begin{equation}
\alpha_{\mathrm{min}}=-k \ln \left\{\frac{k}{\alpha_0}\left(1-e^{-\alpha_0 / k}\right)\right\}  ,
\end{equation}

\noindent where $\alpha_0$ and $h$ are the same parameter to Eq. \ref{Eq:TR}, $k$ is a scaling constant, and $\alpha_{\mathrm{min}}$ is a phase angle where minimum polarization degree ($P_\textrm{min}$) locates. The best-fit parameters and their 1-sigma uncertainties derived from the fitting using various phase angle ranges are summarized in Table \ref{table:a1}. From the fitting using all observation data, we also derive the minimum polarization degree of $P_\mathrm{min}=-0.96^{+0.32}_{-0.34}$ \% at the phase angle of $\alpha_\mathrm{min}=10.7 \degr$  $^{+2.2 \degr}_{-2.8 \degr}$ from TF and $P_{\mathrm{min}}=1.46\pm0.23$ \% and $\alpha_{\mathrm{min}}=8.9 \degr$ $^{+0.4 \degr}_{-0.5 \degr}$ from LE function.

The polarimetric slope, $h$, is known to have a strong correlation to the geometric albedo in the $V$ band ($p_V$) \citep{1967AnWiD..27..109W}. The relation is given as

\begin{equation}
    \log_{10} (p_V) = C_1 \log_{10} (h) + C_2~,
    \label{Eq:8}
\end{equation}

\noindent where $C_1$ and $C_2$ are constant values: $C_1=-1.111\pm0.031$ and $C_2=-1.781\pm0.025$ in \citet{Cellino2015}, and $C_1=-1.016\pm0.010$ and $C_2=-1.719\pm0.012$ in \citet{Lupishko2018}. We assumed this relation and constant parameters are applicable to $R_\mathrm{C}$-band data and derive the geometric albedo as shown in Table \ref{table:3}. We converted the $R_\mathrm{C}$-band geometric albedo ($p_{R_\mathrm{C}}$) into the standard $V$-band geometric albedo ($p_V$) using the color index:

\begin{equation}
    p_V=p_{R_\mathrm{C}} \times 10^{0.4 \left[(V-R_\mathrm{C})_{\odot} - (V-R_\mathrm{C})_{\textrm{WH}}\right]}~,
\end{equation}

\noindent where $(V-R_\mathrm{C})_{\odot} = 0.356 \pm 0.003$ \citep{Ramirez2012} and $(V-R_\mathrm{C})_{\mathrm{WH}} = 0.378\pm0.025$ \citep[converted from a $r' - i'$ color using relations from][]{Jester2005} are $V-R_\mathrm{C}$ colors of the Sun and WH, respectively \citep{Urakawa2011}. The estimated albedo values using $h$ values from the TF fitting are $0.094\pm0.018$ with coefficients in \citet{Cellino2015} and $0.093\pm0.015$ with those in \citet{Lupishko2018}, while those from the LE relation fitting are $0.076\pm0.010$ and $0.077\pm0.008$ with coefficients from \citet{Cellino2015} and \citet{Lupishko2018}, respectively. Therefore, it is safe to say that the possible range of $p_V$ of WH is 0.066--0.112. These results are also summarized in Table \ref{table:3}, together with results from data with different $\alpha$ ranges used for the fitting.

\begin{table*}
\caption{Summary of albedo derivation}
\label{table:3}
\centering
\begin{tabular}{cccccc}
\hline\hline
Function & $\alpha$ range (N\tablefootmark{a}) & $h$ & Reference\tablefootmark{b} & $p_{R_c}$ & $p_V$ \\ \hline
\multicolumn{1}{c}{\multirow{6}{*}{Trigonometric}} & \multicolumn{1}{c}{\multirow{2}{*}{All (19)}} & \multicolumn{1}{c}{\multirow{2}{*}{$0.206\pm0.032$}} & C15 \tablefootmark{c} & $0.096\pm0.018$ & $0.094\pm0.018$ \\
\multicolumn{1}{c}{} &  &  & L18 \tablefootmark{d}  & $0.095\pm0.015$ & $0.093\pm0.015$ \\  
\multicolumn{1}{c}{} & \multicolumn{1}{c}{\multirow{2}{*}{$ < 50\degr$ (7)}} & \multicolumn{1}{c}{\multirow{2}{*}{$0.207\pm0.050$}} & C15 & $0.095\pm0.027$ & $0.093\pm0.026$ \\ 
\multicolumn{1}{c}{} &  &  & L18 & $0.095\pm0.023$ & $0.093\pm0.023$ \\ 
\multicolumn{1}{c}{} & \multicolumn{1}{c}{\multirow{2}{*}{$ < 30\degr$ (4)}} & \multicolumn{1}{c}{\multirow{2}{*}{$0.287\pm0.143$}} & C15 & $0.066\pm0.037$ & $0.065\pm0.036$ \\
\multicolumn{1}{c}{} &  &  & L18 & $0.068\pm0.034$ & $0.067\pm0.034$ \\ \hline
\multicolumn{1}{c}{\multirow{6}{*}{Linear-Exponential}} & \multicolumn{1}{c}{\multirow{2}{*}{All (19)}} & \multicolumn{1}{c}{\multirow{2}{*}{$0.248\pm0.025$}} & C15 & $0.078\pm0.010$ & $0.076\pm0.010$ \\ 
\multicolumn{1}{c}{} &  &  & L18 & $0.079\pm0.008$ & $0.077\pm0.008$ \\
\multicolumn{1}{c}{} & \multicolumn{1}{c}{\multirow{2}{*}{$ < 50\degr$ (7)}} & \multicolumn{1}{c}{\multirow{2}{*}{$0.220\pm0.030$}} & C15 & $0.089\pm0.015$ & $0.087\pm0.015$ \\
\multicolumn{1}{c}{} &  &  & L18 & $0.089\pm0.013$ & $0.087\pm0.013$ \\
\multicolumn{1}{c}{} & \multicolumn{1}{c}{\multirow{2}{*}{$ < 30\degr$ (4)}} & \multicolumn{1}{c}{\multirow{2}{*}{$0.197\pm0.046$}} & C15 & $0.101\pm0.027$ & $0.099\pm0.027$ \\
\multicolumn{1}{c}{} &  &  & L18 & $0.099\pm0.024$ & $0.098\pm0.023$ \\ \hline
\end{tabular}
\tablefoot{
    \tablefoottext{a}{Numbers of data used for fitting}
    \tablefoottext{b}{Reference for coefficients $C_1$ and $C_2$ in Eq. \ref{Eq:8}}
    \tablefoottext{c}{\citet{Cellino2015}}
    \tablefoottext{d}{\citet{Lupishko2018}}
    }
\end{table*}

\subsection{Photometric results} \label{subsec:pho_res}
From the photometry, we find that the nuclear magnitudes in 1949 are $14.86\pm0.20$ in the $B$ band and $14.05\pm0.20$ in the $R_\mathrm{C}$ band. We compared our $R_\mathrm{C}$-band magnitude of the nucleus to the phase function given in \citet{Ishiguro2011} using the heliocentric distance $r_\mathrm{h}=1.148$ au, the geocentric distance $\Delta=0.025$ au, and the phase angle $\alpha=40.1 \degr$, where WH was active in 1949. From the phase function, the expected $R_\mathrm{C}$-band reduced magnitude at the time of 1949 observation is $14.4\pm0.1$. We find that this $R_\mathrm{C}$-band reduced magnitude is $\sim 0.4$ magnitude brighter than the expected nuclear brightness based on the phase function of the WH's nucleus \citep{Ishiguro2011}. Considering the magnitude error (0.2), we conclude that WH's nucleus was brightened when it showed the comet-like tail.

In addition, we find $B-R_\mathrm{C}$ of the nucleus in 1949 was $0.80 \pm 0.29$, consistent with previous reports by \citet{Lowry2003} ($0.81 \pm 0.06$) and \citet{Urakawa2011} \citep[$0.80 \pm 0.04$, with the magnitude converted from SDSS g', r' and i' to Johnson-Cousins system using][]{Jester2005}  when it was inactive.
We also derive the total signal within the tail region in the $B$ band and the $R_\mathrm{C}$ band to be $(1.04 \pm 0.09) \times 10^{-15}$ W m$^{-2}$ and $(1.78 \pm 0.09) \times 10^{-15}$ W m$^{-2}$, respectively. The corresponding surface brightnesses of the tail are $25.96\pm0.22$ in the blue plate and $25.44\pm0.21$ magnitudes arcseconds$^{-2}$ in the red plate. We confirm that the $B-R_\mathrm{C}=0.52\pm0.30$ color of the tail is consistent with the nucleus within a 1-sigma confidence interval. Although the tail color is bluer than the Sun, the similarity in color between the tail and the nucleus may suggest that the tail consists of dust grains with a similar optical color to the nucleus, and was detectable via the scattered sunlight.

    \begin{figure}
    \centering
    \includegraphics[width=\hsize]{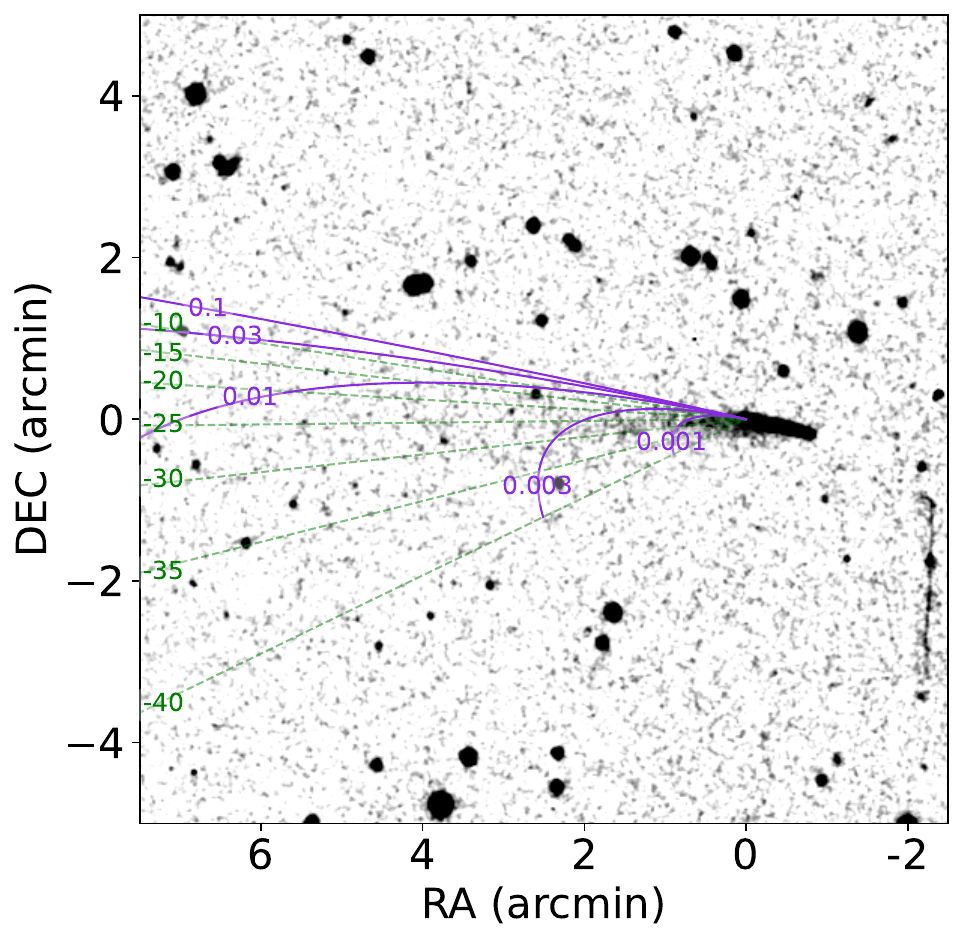}
        \caption{
        Synchrone (green dotted lines) and syndyne (violet solid lines) network drawn on the blue plate image on November 19, 1949. This image has a standard orientation, that is, north is up and east to the left. The numbers on the synchrone curves correspond to days of dust ejection before the observed day, while the numbers on the syndyne curves indicate the corresponding $\beta$ values. 
        }
        \label{Fig5}
   \end{figure} 

    \begin{figure}
    \centering
    \includegraphics[width=\hsize]{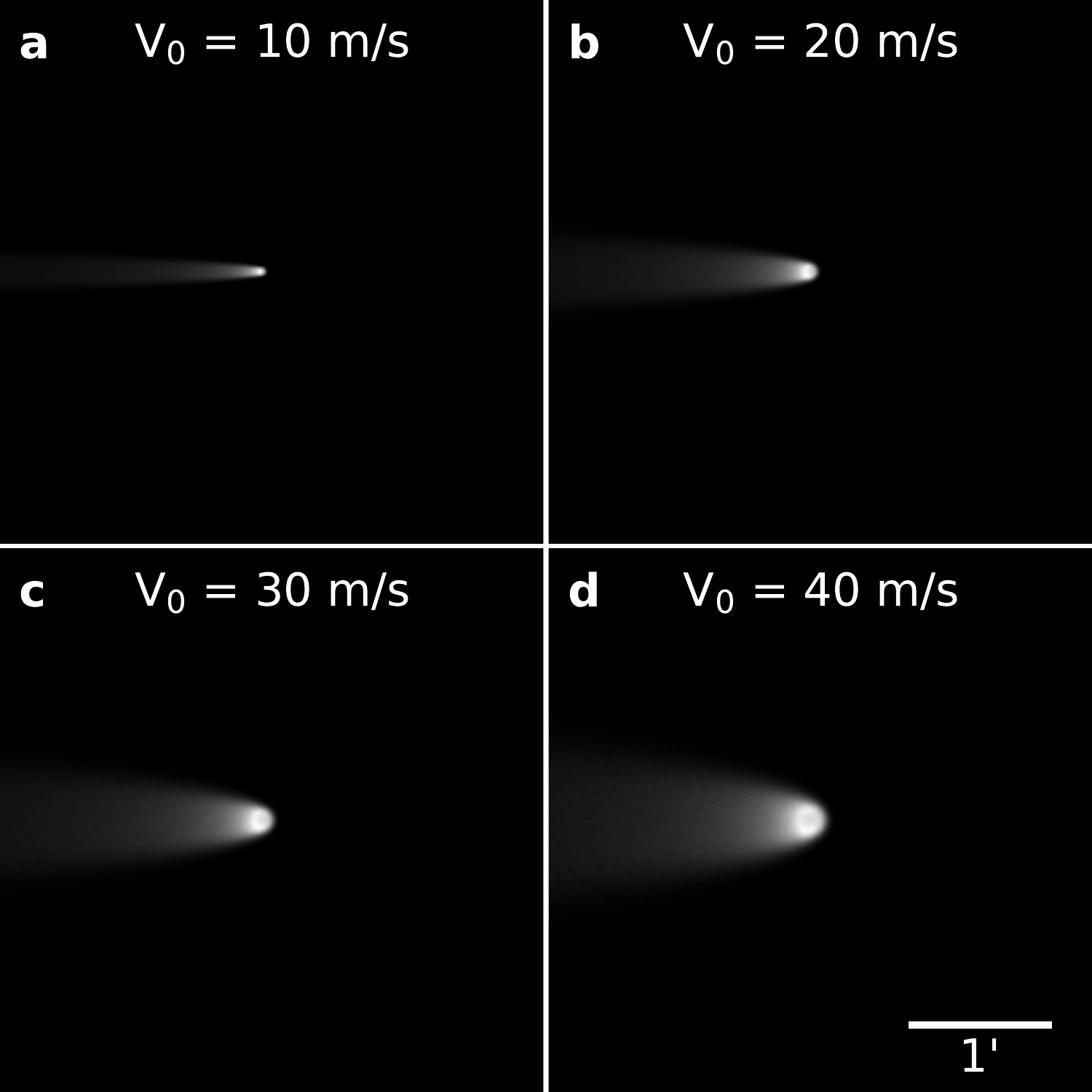}
        \caption{Simulated images of WH with different $V_0$. We assumed that dust ejection happened for one day impulsively around 29.5 days before the observation. The field-of-view of each image is identical to the one of Fig. \ref{Fig1} (b). 
        The comet-like ejection velocity, $V(a)=V_0 \times \sqrt{\beta}$, was used for this model. We assumed hemispherical dust ejection with fixed $\beta$ range ($3\times 10^{-4} \leq \beta \leq 1\times 10^{-2}$) and $q$ for the differential size distribution.}
        \label{Fig6}
   \end{figure}

\subsection{Results of dust tail analysis} \label{subsec:tail_res}

As shown in Sect. \ref{subsec:pho_res}, the color of the tail is similar to the Sun and the nucleus. We also find that the nuclear magnitude in the 1949 images was brighter than the expected magnitude of the nucleus. This evidence suggests that WH ejected dust particles around the time of the 1949 observation, and the associated dust tail was imaged. 

Based on the idea above, we investigated the ejection epoch and size of the dust particles from the tail morphology (see Sect. \ref{subsec:tailmodel} for model description). Fig. \ref{Fig5} compares the observed images with a synchrone-syndyne network. Although the synchrone-syndyne analysis is a simple model that assumes a zero initial velocity, this analysis derives a rough estimate of the particle size and the ejection time. The position angle of the observed tail is $105.5\degr\pm 6.5\degr$. The best-match synchrone curve corresponds to 25--32 days before the observed date (November 19, 1949). In other words, if the ejection velocity was small, the dust particles were likely ejected within 6 days on October 18--25, 1949. This period of activity is likely shorter than 7 days in the realistic case (i.e., non-zero ejection velocity). Moreover, the intersection of the synchrone and syndyne curves indicates $\beta$ is ranging from 
$\sim 3\times10^{-4}$ to $\sim 1\times10^{-2}$, which correspond to $a$=44 \textmu m --1.5 mm when we assume the Ryugu-like dust mass density \citep[$\rho_\mathrm{d}$=1\,300 kg m$^{-3}$, ][]{Miyazaki2023}. This maximum radius ($a$=1.5 mm) is a lower limit. It would be possible that larger dust particles might be present near the nucleus and enhance the nuclear magnitude.

Figure \ref{Fig6} shows the dust simulation results, where we take into account the dust ejection velocity. Since our synchrone analysis suggests that the ejection duration was short, for only 7 days or less, we considered an impulsive dust ejection. In this model, we assumed that dust ejection happened for only $\Delta t=1$ day on October 29.5, 1949 (the middle day from our synchrone analysis) from the entire Sun-lit hemisphere. We also assumed a power index $q = -3.5$ for the differential size distribution of the dust particles. These parameters ($\Delta t$ and $q$) cannot be determined from the observed images not only because of the faintness of the tail but also because of the lack of time-series images. In Fig. \ref{Fig6}, four different terminal velocities ($V_0$=10, 20, 30, and 40 m s$^{-1}$) were tested. By visual inspection, the model image with
$V_0<$20 m s$^{-1}$ matches the observation. The visible region of the tail has a width of $\approx$8 \arcsec, which matches the model image with $V_0 \sim $10 m s$^{-1}$. With Eq. \ref{Eq:V0} and $V_0\sim$10 m s$^{-1}$, our model suggests a terminal ejection velocity of 1 mm-sized dust particles ($\beta=4.4\times10^{-4}$) would be 0.2 m s$^{-1}$. This velocity is even slower than the escape velocity from WH (a bulk density of 1\,190 kg m$^3$ and a diameter of $\sim$4\,000 m are assumed \citep{Watanabe2019,Bach2017}).

 Finally, we derived the dust mass. We obtained the flux ratio of the dust tail with respect to the nucleus as $\frac{I_\mathrm{tail}}{I_\mathrm{WH}} = 0.139$ from the blue plate. Substituting maximum (1.5 mm) and minimum (44 \textmu m) dust sizes and assuming $q=-3.5$, we find the dust total mass in the tail was $9.27 \times 10^5$ kg using Eqs. (\ref{Eq:mass})--(\ref{Eq:No}). The estimated dust mass is only $1.8 \times 10^{-6} $ \% of WH nucleus mass (when we assume that the bulk density of WH is 1\,190 kg m$^3$, similar to another C-complex asteroid (162173) Ryugu \citep{Watanabe2019}. We note that this estimated dust mass would be a lower limit since large particles residing near the nucleus were not included in the calculation.

\section{Discussion}\label{sec:discussion}
We have investigated the WH's origin and its mysterious activity in 1949 from multiple perspectives based on photometry, spectroscopy, and dust dynamical properties. As we mentioned in Sect. \ref{sec:intro}, the difference between comets and asteroids has become unclear nowadays because of recent findings of objects having the hybrid natures of comets and asteroids \citep{Jewitt2022}. However, we adopt a conventional classification based on the origins in the following discussion. We thus refer to a comet as an object originating from the trans-Neptunian region \citep{1997Icar..127...13L}, while an asteroid is an object in or from the main belt. In the following subsections, we discuss each feature and consider the possible mechanism for the 1949 activity.

\subsection{The optical nature}\label{subsec:nature}
First, we discuss the optical nature. We derived the colors of the nucleus $B-R_\mathrm{C} = 0.80\pm0.29$ and the tail $B-R_\mathrm{C} = 0.52\pm0.30$. Although these errors are not small enough to discuss the origins, it is safe to say that the color of the tail is consistent with the nucleus. From this consistency, we conclude that the tail is dust-scattered light. The 0.4 magnitude enhancement of the nucleus can be explained by large dust particles distributed around the Hill radius.

From previous research, it is known that the nucleus color is slightly bluer than that of the Sun \citep{Chamberlin1996}.  In general, comets exhibit red spectral characteristics compared to the solar spectrum \citep{2002AJ....123.1039J}. However, the blue color of the nucleus is not necessarily a condition to reject the possibility that WH is a cometary origin. In fact, some comets (e.g., 95P/Chiron) indicated blue  properties \citep{1993Icar..104..138L,1999A&A...341..912B}. More comprehensive spectroscopic research in the optical and near-infrared wavelengths suggested that the WH spectrum is in agreement with carbonaceous chondrite meteorites \citep{Kareta2023}, which originate from C-type asteroids. Accordingly, from the reflectance color and spectrum, WH likely originated from the main belt. 

\begin{figure}
    \centering
    \includegraphics[width=\hsize]{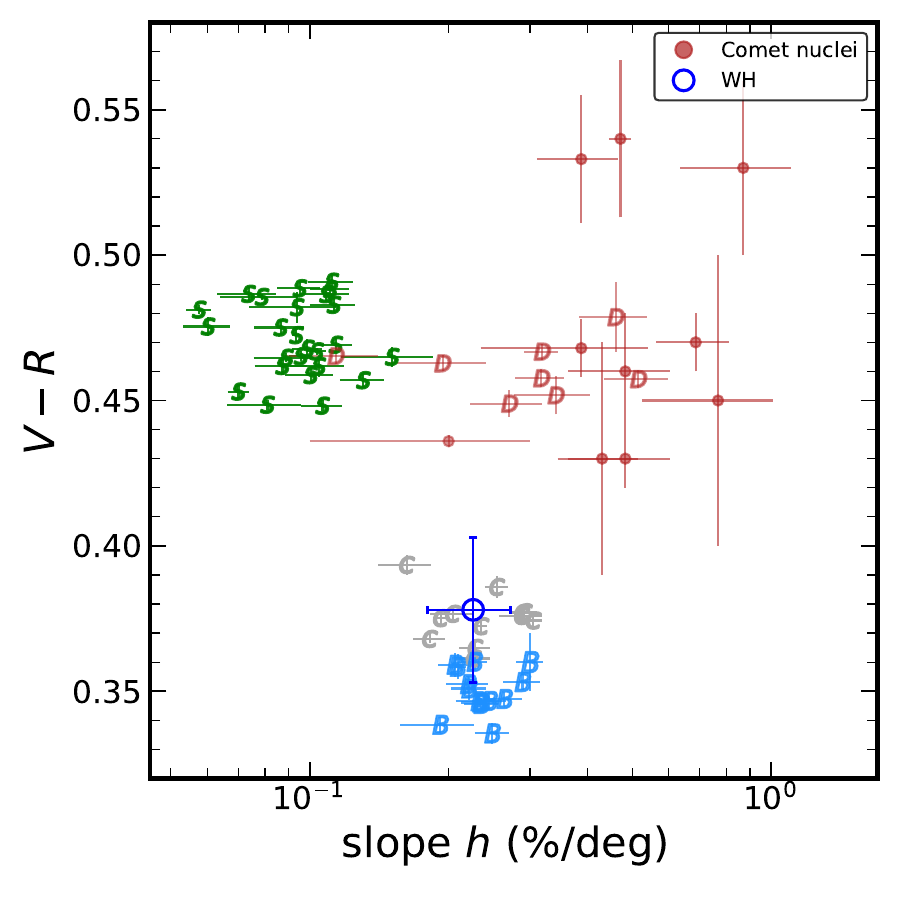}
        \caption{
        Polarimetric slope $h$ and $V-R_\mathrm{C}$ plot of WH in comparison with different types of asteroids and comet nuclei. The comparison data are taken from \citet{Geem2022}. Points with S, C, B, and D denote S-, C-, B-, and D-type asteroids.
              }
        \label{Fig7}
   \end{figure}

We also investigated polarization properties. No polarimetric information was reported before our work. Based on our data, we estimated the range of geometric albedo ($p_\mathrm{V}$) to be 0.066 -- 0.112. The derived geometric albedo is marginally consistent with or slightly higher than the previous estimates: $p_\mathrm{V}=0.055\pm0.012$ \citep{Ishiguro2011}, $0.040-0.055$ \citep{Bach2017}, and $0.059\pm0.011$ \citep{2009A&A...507.1667L}. This slight discrepancy may be due to a slight underestimation of the inherent errors in the models used in each study (including our research).  \citet{Geem2022} proposed a technique for discriminating between comet-like and asteroid-like objects by combining a polarimetric slope ($h$) and color index ($V-R_\mathrm{C}$).  In Fig. \ref{Fig7}, we applied the method in \citet{Geem2022} to our target. Indeed, WH is consistent with C- and B-type asteroids. However, the albedo of comets and some dark asteroids (C-complex and D-types) are largely overlapped with each other \citep{Lamy2004,2011PASJ...63.1117U,2014ApJ...792...30M}, implying the geometric albedo tightly related to the polarimetric slope \citep[$h$ in][]{Geem2022} cannot be a 'conclusive' determinant. Also, as noted above, some exceptional cometary samples have unusually blue color indices. In other words, although the method of \citet{Geem2022} is a useful tool that can make rough discrimination of comets and asteroids, it is not necessarily possible to conclude the origin with only $h$ and $V-R_\mathrm{C}$. Further careful consideration is needed, as given in Sect. \ref{subsec:mechanism}.

Furthermore, we show that the WH's large polarization degrees at large phase angles ($\alpha>40 \degr$). These values are higher than those of the nucleus of a nearly dormant comet 209P/LINEAR. It is, however, true that the number of polarization observation samples of cometary nuclei at large phase angles is insufficient because the nuclei should be enclosed by thick dust comae at the large phase angles (i.e., the small heliocentric distances). In addition, \citet{Kwon2018} speculated that large dust might have accumulated on the old comet nuclei, and the effect is responsible for large polarization degrees at high phase angles. Indeed, WH has been in a stable orbit for the past 4\,000 years or more, maintaining a solar distance of $\sim1$ au \citep{Kareta2023}. Therefore, it may be possible that the surface materials are thermally metamorphosed or sorted to extract only large dust particles.

Summarizing the discussion regarding the optical nature, WH has a fairly high probability of being of main-belt origin. This fact is also in accordance with dynamical studies \citep{Bottke2002,Granvik2018}, although dynamical research has an intrinsic difficulty due to the chaotic orbit of near-Earth objects.

\subsection{The cause of the dust ejection in 1949}\label{subsec:mechanism}
We subsequently discuss the essential question of why WH exhibited comet-like activity in 1949 based on the abovementioned fact that WH is likely the main belt origin. \citet{Fernandez1997} pointed out that the color of the tail was remarkably blue. Our analysis shows that the color of the tail is consistent with the color of the nucleus within the error bar. Actually, our intensity estimate of the tail ($1.20 \times 10^{-15}$ W m$^{-2}$ and $1.86 \times 10^{-15}$ W m$^{-2}$) is consistent with \citet{Fernandez1997} within a factor of 2. The difference would be caused by background detection since the signal of the tail is very weak compared to the background. For reference, we recorded our method for photometric analysis in Appendix \ref{sec:A1}. Since the details of the tail color analysis are not provided in \citet{Bowell1992} (where the authors describe the very blue color), it is difficult to compare our results to their report, and we thus proceed with the discussion based on our findings.

Because of the color similarity between the tail and the Sun, we assume that the tail was responsible for the scattered sunlight by the ejected dust particles. Under this assumption, we investigated the particle size, total mass, and epoch/duration of dust ejection. From synchrone--syndyne analysis, we found that the particle size is significantly larger than the wavelength, having the size parameter $X=\frac{2 \pi a}{\lambda}\gtrsim 100$. This large dust size is also consistent with the neutral tail color, which is similar to sunlight.

After the publication of \citet{Fernandez1997}, a new concept of active asteroids emerged. Asteroids eject their mass through lofting via electrostatic repulsion or thermal radiation pressure, thermal fatigue, impact, ice sublimation, and rotational disruption \citep{Jewitt2012,2021A&A...654A.113B,Molaro2020}. Various recent observations widely support these ideas. Therefore, it is natural to assume that WH is also an active asteroid of main-belt origin, and it indicated an activity in 1949 for a reason. 

First, we consider the possibility of dust lofting by either electrostatic repulsion or thermal radiation pressure. \citet{Jewitt2012} estimated dust size that can be ejected from a 1 km asteroid around 1 au as 1.5-5 \textmu m, suggesting that only $\lesssim 1$ \textmu m-sized dust can be lifted on the WH's surface by electrostatic repulsion. In addition, \citet{2021A&A...654A.113B} examined dust lofting via thermal radiation pressure from the heated surface with various sizes at various solar distances. For the WH's size asteroid at its perihelion, no dust particles can be lifted up according to their calculation. Therefore, the hypothesis of dust lofting via these mechanisms can be ruled out since the particle size from our estimation is larger by at least an order of magnitude. 

Second, thermal fatigue can break surface boulders and trigger exfoliation to launch some fragments. This mechanism was suggested as one of the most probable causes of the (101955) Bennu's activity \citep{Molaro2020}. However, WH was active only once in 1949, although there were $> 15$ observational opportunities (i.e., the perihelion passages) since 1949. Furthermore, it is difficult to explain WH activity in terms of the mass of the ejecta by thermal fatigue. We estimated that $9.27\times10^5$ kg of dust was ejected from the WH within about one day. This is equivalent to a dust ejection rate of 10.7 kg s$^{-1}$. Since the mass ejection rate observed at Bennu is 10$^{-7}$ kg s$^{-1}$ \citep{2020JGRE..12506381H}, WH ejected nearly 8 orders of magnitude more dust than Bennu. The short duration of the ejection is also quite different from the nature of dust ejections observed at Bennu. Therefore, thermal fatigue is a mechanism less likely to cause the WH's 1949 activity. 

Third, an impact may trigger the activity of an asteroid \citep[see, e.g.,][]{2011ApJ...740L..11I}. We estimated the diameter of a hypothetical impactor, which can produce ejecta whose mass is equivalent to tail mass, using a model in \citet{Holsapple2022}. We find that a $\sim 1$m-sized object should collide with WH to generate the ejected mass. From interplanetary dust particle impact flux by \citet{Grun1985} and surface area estimated from a diameter by \citet{Bach2017}, such impact would occur with a timescale of $10^8$ years, longer than a typical lifetime of near-Earth asteroids. Although the impact probability is not zero (once per $10^8$ years), the possibility of impact-triggered activity is considerably low.

Fourth, we consider volatile sublimation as observed in main-belt comets \citep[MBCs,][]{2006Sci...312..561H}. The activities of MBC samples due to ice sublimation usually continued for several months \citep{2011AJ....142...29H,2011ApJ...736L..18H,2014AJ....147..117J}. The dust ejection velocity from MBCs is typically close to the escape velocity \citep{2014AJ....147..117J}, which is similar to the WH's 1949 event. Although we derived the duration of the dust ejection of < 7 days from our non-zero ejection velocity model, this estimate should be significantly overestimated because the zero ejection velocity is unrealistic. Adopting the non-zero ejection velocity model, the duration of the activity would be $\sim 1$ days. Considering such an impulsive dust ejection, the possibility of volatile sublimation for the 1949 activity is less likely.

The remaining possibility is that the WH tail was formed by rotational instability \citep[see e.g.,][]{2013ApJ...778L..21J}. The low ejection velocity close to the escape velocity and the presence of large particles are consistent with the mechanism. We derived the total dust mass of $9.27 \times 10^5$ kg, which is only $\sim10^{-6}$ \% of the nuclear mass ($\sim 5\times10^{13}$ kg). Such rotation-induced mass shedding has been observed in various asteroids \citep{2009AJ....137.4296J}. In the case of mass shedding, because the centrifugal force accelerates the surface material, large particles can also be ejected with nearly an escape velocity. Therefore, rotation-induced mass shedding is the most likely scenario for dust ejection based on our tail analysis. 

Our inference can be supported by the light curve analyses \citep{Harris1983, Osip1995, Urakawa2011}. The rotation period varies widely from 3.556 to 7.15 hours in the literature. \citet{Harris1983} reported a rotation period of 3.556 hours, while \citet{Osip1995} reported $6.1 \pm 0.2$. \citet{Urakawa2011} conducted a comprehensive analysis of the shape model, assuming the derived rotational period of 7.15 hours. However, \citet{Urakawa2011} also suggested the possibility of the solution for 3.58 hours. If the shorter solution is correct, WH still maintains a rapid rotation period. Such fast-rotating asteroids may create top shape, as observed for, for example, (66391) Moshup, (101955) Bennu, (162173) Ryugu  \citep{Ostro2006, Lauretta2019, Watanabe2019}, and the shape reduces the rotational variability of the magnitude because of the axis-symmetrical shape.

In summary, WH is very likely an active asteroid of main belt origin. Although the possibility of volatile sublimation cannot be ruled out from our analysis, the possibility of rotational mass shedding is most likely a scenario for the 1949 activity based on our data and analyses, as well as the light curve properties derived by other research groups.

\begin{acknowledgements}
We deeply appreciate an anonymous reviewer and Dr. Emmanuel Lellouch for the valuable comments and suggestions. This research at SNU was supported by a National Research Foundation of Korea (NRF) grant funded by the Korean government (MEST) (No. 2023R1A2C1006180). The Pirka, Nayuta, and Kanata telescopes are operated by the Graduate School of Science, Hokkaido University; the Center for Astronomy, University of Hyogo; and the Hiroshima Astrophysical Science Center, Hiroshima University, respectively. These telescopes are partially supported by the Optical and Infrared Synergetic Telescopes for Education and Research (OISTER) program funded by the MEXT of Japan. SI acknowledges financial support from INAF - Call for fundamental research 2022, Minigrant RSN3.
\end{acknowledgements}

\bibliographystyle{aa}
\bibliography{ref_WH}
%
%

\begin{appendix} 
\section{Linearization of photographic plate images}\label{sec:A1}
For the linearization of signals from photographic plates, we utilized an empirical relation from \citet{Cutri1993}, originally introduced by \citet{DeV1968},

\begin{equation}
    I = A \omega ^ n.
\label{Eq_a_I}
\end{equation}

\textit{I} is an intensity of incident light, \textit{$\omega$} is an opacitance of the emulsion on the photographic plate, and \textit{A} and \textit{n} are constants. The opacitance can be calculated from the density of the emulsion $\delta$ as below:

\begin{equation}
    \omega = 10^\delta-1.
\end{equation}

Accordingly, the opacitance of the region with a zero emulsion density becomes 0. Then, the magnitude of a point source from aperture photometry (\textit{m}) is given by

\begin{eqnarray}
    m & = & -2.5 \textrm{log}_{10} \int I_{\textrm{source}} d\Omega \\
    & = & -2.5 \textrm{log}_{10} \int \left(I_{\textrm{aper}} - I_{\textrm{sky}}\right) d\Omega \\
    & = & -2.5 \textrm{log}_{10} A - 2.5 \textrm{log}_{10} \int \left(\omega_{\textrm{aper}} ^ n - \omega_{\textrm{sky}} ^ n\right) ,
\end{eqnarray}

\noindent where $I_{\textrm{source}}$ is the intensity from the source, and $I_{\textrm{aper}}$, $I_{\textrm{sky}}$, $\omega_{\textrm{aper}} ^ n$ and $\omega_{\textrm{sky}} ^ n$ are intensities and opacitances within the aperture and the sky background, respectively.

We first set $n$ in Eq. A.1 to 1 as an initial guess and conducted aperture photometry on the image using an aperture with the size of $\sim 2 \times$ full-width half maximum of the point spread function in the image. Then, we compared the magnitudes of stars from plates to those of equivalent stars at the Pan-STARRS DR1 catalog, whose location is consistent within 3 \arcsec difference. \citep{Chambers2016}. We converted Pan-STARRS \textit{g},\textit{r}, and \textit{i}-band magnitudes into Johnson $B$- and Cousins $R_\mathrm{C}$-band using the relation from \citet{Kostov2017} to derive catalog magnitudes ($m_{\textrm{cat}}$). \citet{Cutri1993} showed the consistency between signals measured from Blue and Red plate images and those from photometry on CCD images with Johnson $B$- and Cousins $R_\mathrm{C}$-band, and their difference is up to 0.3 magnitudes. For the blue plate image, we used 79 stars whose magnitude ranges from 17 to 19 magnitudes. Meanwhile, we used 120 stars with a magnitude range of 16 -- 18 magnitudes for the red plate. Then, we plotted the catalog magnitude ($m_{\textrm{cat}}$) derived above on the x-axis and the plate magnitude ($m_{\textrm{plate}}$) on the y-axis, defined as

\begin{equation}
    m_{\textrm{plate}} = -2.5 \textrm{log}_{10} \int \left(\omega_{\textrm{aper}} ^n - \omega_{\textrm{sky}} ^ n\right). 
\end{equation}

We calculated the slopes $a(n)$ and their intercepts $b(n)$ of a linear regression line using the least square approximation. We determined the $n$ value for each plate that makes the slope to be unity using the chi-square minimization technique. We calculated the reduced chi-square value using the formula below,

\begin{equation}
    \chi^2_{\textrm{red}} = \sum^{N}_{i = 1} \left(\frac{m_{\textrm{cat}, i}-m_{\textrm{plate}, i} - b}{\sigma_{i}}\right)^2
\end{equation}

\noindent where $m_{\mathrm{cat}, i}$ and $m_{\mathrm{plate}, i}$ are catalog and plate magnitudes of the $i$-th star within the image. 1-sigma confidence intervals of $n$ is estimated by $\chi^2_{\textrm{red}}(n) < \chi^2_{\textrm{red, min}}(n)+\sqrt{\frac{1}{N-2}}$.

We derived the $A$ values using the formula in \citet{Cutri1993},

\begin{equation}
    A = 10^{\frac{b}{-2.5}} I_0 \Delta \lambda_{\textrm{eff}},
\end{equation}

\noindent where $I_0$ is zero magnitude intensity of a filter, and $\Delta \lambda_{\textrm{eff}}$ is its effective width of the filter. We used zero magnitude intensities and effective widths of Johnson $B$ and Cousins $R_\mathrm{C}$ from the Spanish Virtual Observatory (SVO) Filter Profile Service \footnote{http://svo2.cab.inta-csic.es/theory/fps/}. 1-sigma confidence intervals of the $A$ values are calculated from those of $n$. Finally, the linearization formulae derived for blue and red plate images are shown below:

\begin{eqnarray}
    I_{\textrm{Blue}} = (9.972\pm3.682) \times 10^{-16} \times \omega ^{0.594\pm0.066}~,  \\
    I_{\textrm{Red}} = (7.564\pm1.183) \times 10^{-15} \times \omega ^{0.951\pm0.029}~.
\end{eqnarray}

\section{Error estimation of photographic plate images}\label{sec:A2}
In this study, we considered three sources of uncertainty regarding digitized photographic plate images. The first uncertainty source is from the linearization coefficients ($\sigma_{A}, \sigma_{n}$) we derived in Sect. \ref{sec:A1}. The second one is the inhomogeneous thickness of the emulsion, which causes background fluctuation in the image. The level of background fluctuation is estimated by background RMS value ($\sigma_{\textrm{bg, RMS}}$) calculated by Source-Extractor \citep{Bertin1996}. The last one is the Poisson noise of incident photons, $\sigma_{\textrm{Poisson}} = \sqrt{N}$, where $N$ is the number of incident photons to the plate. We estimated $N$ from intensities using the following equation

\begin{equation}
    N = \frac{It_{\textrm{exp}}}{h\nu_{\textrm{eff}}}~,
\end{equation}

\noindent where $t_{\textrm{exp}}$, \textit{h}, and $\nu_{\textrm{eff}}$ are the exposure time, Planck constant, and the effective frequency of each band, respectively.
From Eq. \ref{Eq_a_I}, the Poisson noise of $\omega$ ($\sigma_{\omega, \textrm{Poisson}}$) is given as

\begin{equation}
    \sigma_{\omega, \textrm{Poisson}} = \frac{\omega}{n\sqrt{N}}.
\end{equation}

With $\sigma_{A}$, $\sigma_{n}$, and $\sigma_{\omega, \textrm{Poisson}}$, we can derive the RMS values of $\omega^{n}$ and $I$ using the following equations

\begin{eqnarray}
    \sigma_{\omega^{n}} &=& \omega^{n} \sqrt{\left(\frac{n}{\omega}\sigma_{\omega,  \textrm{Poisson}}\right)^2+(\mathrm{ln} \ \omega \ \sigma_{n})^2}~,  \\
    \sigma_I &=& I \sqrt{\left(\frac{\sigma_A}{A}\right)^2+\left(\frac{\sigma_{\omega^{n}}}{\omega^{n}}\right)^2}~.
\end{eqnarray}

Finally, an uncertainty of a background-subtracted signal ($I_{\textrm{sig}} = I - I_{\textrm{bg}}$), $\sigma_{I_{\textrm{sig}}}$, is given as

\begin{equation}
    \sigma_{I_{\textrm{sig}}} = \sqrt{\sigma_I^2+\sigma_{I_{\textrm{bg}}}^2+\sigma_{I_{\textrm{bg, RMS}}}^2} .
\end{equation}

\section{Result of fitting for PPCs}
The results from the fitting of PPCs are summarized in Table \ref{table:a1}.

\begin{table*}
\caption{Summary of best-fit parameters and their 1-sigma uncertainties.}
\label{table:a1}
\centering
\begin{tabular}{ccccccc}
\hline\hline
Function & $\alpha$ range (N\tablefootmark{a}) & $\alpha_0$ (deg) & $h$ (\% deg$^{-1}$) & $c_1$ & $c_2$ & $k$  \\ \hline
\multicolumn{1}{c}{\multirow{3}{*}{Trigonometric}} & All (19) & $18.750\pm1.146$ & $0.208\pm0.030$ & $0.848\pm0.139$ & $-0.657\pm0.231$ &  \\
\multicolumn{1}{c}{} &  $ < 50\degr$ (7) & $19.720\pm1.540$ & $0.207\pm0.050$ & $1.221\pm0.500$ & $-0.343\pm5.527$ &  \\  
\multicolumn{1}{c}{} &  $ < 30\degr$ (4) & $20.882\pm2.567$ & $0.287\pm0.143$ & $1.902\pm1.386$ & $0.203\pm5.813$ & \\ \hline
\multicolumn{1}{c}{\multirow{3}{*}{Linear-Exponential}} & All (19) & $18.858\pm0.990$ & $0.248\pm0.025$ &  &  & $21.872\pm2.704$ \\
\multicolumn{1}{c}{} &  $ < 50\degr$ (7) & $19.605\pm1.224$ & $0.220\pm0.030$ &  &  & $279.015\pm132.123$ \\ 
\multicolumn{1}{c}{} &  $ < 30\degr$ (4) & $19.145\pm1.554$ & $0.197\pm0.046$ &  &  & $251.271\pm144.423$ \\ \hline
\end{tabular}
\tablefoot{
    \tablefoottext{a}{Numbers of data used for fitting}
    }
\end{table*}

\end{appendix}

\end{document}